\documentstyle[preprint,aps]{revtex}

\begin{document}
\draft

\title{Chaotic scattering around black holes}

\author{Juan~M.~Aguirregabiria}
\address{Fisika Teorikoa, Euskal Herriko Unibertsitatea, 
P.~K.~644, 48080 Bilbo, Spain }

\maketitle

 \begin{abstract}
It is shown that the scattering of a charged test particle by a system
of four extreme Reissner-Nordtr\"om black holes is chaotic in some
cases. The fractal structure of the scattering angle and time delay
functions is another manifestation of the existence of a nonattracting
chaotic set: fractal basin boundaries were previously known. 
 \end{abstract}

 \pacs{04.25.-g,04.40.Nr,04.70.Bw,95.10.Fh}

\section{Introduction}
\label{sec:intro}

The appearance of deterministic chaos in general relativity has been 
analyzed mainly in two different contexts: the evolution of 
cosmological models\cite{Hobill} and the motion of test particles in 
prescribed backgrounds. In the last context we may mention the study 
of charged particles in interaction with gravitational 
waves\cite{Varvoglis} and in the Ernst spacetime\cite{Karas} and the 
chaotic motion of spinless\cite{Bombelli} and spinning 
particles\cite{Suzuki} around a Schwarzchild black hole and
a Schwarzchild black hole with quadrupolar and octupolar
contributions of arbitrary strength\cite{Letelier} , but the main 
interest has been deserved by the system of two extreme 
Reissner-Nordstr\"om black holes described by the 
Majumdar-Papapetrou (MP) static solutions\cite{M,P}. 

Chandrasekhar\cite{Chandra} and Contopoulos\cite{Conto} analyzed the
timelike and null geodesics around these two black holes and described
the appearance of chaotic trajectories and their geometric source has
been discussed by Yurtsever\cite{Yurtsever}. Another aspect of chaos,
fractal basin boundaries, have been studied by Dettmann, Frankel and
Cornish\cite{Dettmann} for the case of two and three static black
holes in the MP geometry and by Drake, Dettmann, Frankel and
Cornish\cite{SR} in the context of Special Relativity. 
These fractal boundaries are consequence of
the existence of a nonattracting chaotic set whose dimension
provides a coordinate independent measure of chaos\cite{invariant}. 

The goal of this paper is to show that, for a different range of
parameters, another consequence of the existence of a nonattracting
chaotic set arises in multi-black-hole spacetimes: 
chaotic scattering\cite{Ott}.
We will consider a static system of four extreme Reissner-Nordstr\"om
black holes and a test particle, with $e/m >1$. For certain energy
ranges, the scattering is chaotic both in the relativistic system and
in the Newtonian approximation. To the best of our knowledge this is
the first time that chaotic scattering angle and time delay functions
are described in general relativity\cite{private}.

\section{The equations of motion}
\label{sec:motion}

Majumdar\cite{M} and Papapetrou\cite{P}  independently showed that 
the metric
 \begin{equation}
ds^2=-U^{-2}\,dt^2+U^2\left(dx^2+dy^2+dz^2\right)
 \label{metric}
 \end{equation}
is a static solution of the Einstein-Maxwell equations corresponding 
to the electrostatic potential $A=U^{-1}\,dt$ if the function 
$U(x,y,z)$ satisfies Laplace's equation: 
 \begin{equation}
U_{,xx}+U_{,yy}+U_{,zz}=0.
 \label{Laplace}
 \end{equation}
Hartle and Hawking\cite{Hartle} showed that if one chooses
 \begin{equation}
U=1+\sum_{i=1}^{N}{\frac{m_i}{|{\bf x}-{\bf x}_i|}},
 \label{U}
 \end{equation}
the MP solution represents $N$ extremal ($q_i=m_i$) 
Reissner-Nordstr\"om black holes held at the fixed positions, 
${\bf x}={\bf x}_i$, under the combined action of gravitational 
attraction and electrostatic repulsion. The apparent singularity of 
$U$ at the points ${\bf x}_i$ corresponds to the usual 
coordinate singularity at an event horizon. 

Let us now consider a test particle of mass $m$ and charge $e$ moving 
in the above multi-black-hole geometry. Following Dettmann, Frankel 
and Cornish\cite{Dettmann} we can write its equations of motion in the 
form 
 \begin{eqnarray}
\dot {\bf x} &=& U^{-1}{\bf v},\label{xdot}\\
\dot {\bf v} &=& U^{-2}\left[\left(1+2\, {\bf v}\cdot{\bf v}-
                             \frac{e}{m}\gamma\right) \nabla U
                       -\left({\bf v}\cdot\nabla U\right)
                        {\bf v}\right],\label{vdot}\\
\dot t&=&U\gamma,\\
\gamma &\equiv& \sqrt{1+{\bf v}\cdot{\bf v}},
 \end{eqnarray}
where a dot indicates the derivative with respect to the proper 
time and $(\gamma,{\bf v})$ are the components of the 
four-velocity in an orthonormal frame. The conserved energy per unit 
mass is
 \begin{equation}
E= U^{-1}\left(\gamma-\frac{e}{m}\right).
 \label{energy}
 \end{equation}

\section{Chaotic Scattering}
\label{sec:scattering}

Contopoulos\cite{Conto} analyzed the case of an uncharged particle
($e/m=0$) moving around $N=2$ static black holes and discussed the
role of periodic orbits and the route to chaos through a cascade of
period-doubling bifurcations. Dettmann, Frankel and
Cornish\cite{Dettmann} considered the cases $N=2, 3$ for a particle
with $e/m \le 1$ which was released from rest. Under these
circumstances the gravitational attraction overcomes the electrostatic
repulsion and the particle will fall into one of the black holes or
orbit indefinitely, but will not escape to infinity. Dettmann, Frankel
and Cornish showed that the boundary separating in phase-space the
different asymptotic behaviors is fractal, which indicates the
presence of a nonattracting chaotic set\cite{Ott} whose dimension
provides a coordinate independent measure of chaos\cite{invariant}. 

We have considered the motion of a charged test particle in the plane
$(x,y)$ around a system of four equal Reissner-Nordtr\"om black holes
which are at rest at points $(\pm1,\pm1,0)$ and have $q_i=m_i=1/3$.
The energy~(\ref{energy}) has not exactly the same structure as in
Newtonian mechanics, but we can still get the same kind of qualitative
information given by the turning points of the potential if we study
the curves of zero velocity\cite{Conto}, i.e., curves in the $(x,y)$
plane which satisfy Eq.~(\ref{energy}) for ${\bf v} =0$. Orbits never
cross the curves of zero velocity. 

If $e/m>1$ the electrostatic repulsion acting on a stationary test
particle is stronger than the gravitational attraction and for
$(1+2\sqrt{2}/3)^{-1}(1-e/m) < E < 0$ there exist a curve of zero
velocity around each black hole. In Fig.~\ref{fig:0} we have the right
hand side of~(\ref{energy}) for ${\bf v}=0$ and $e/m=2$. The curves of
zero velocity are the contours of constant height of the displayed
surface. In these circumstances, we will have scattering: orbits
coming from infinity will escape to infinity, they will not fall into
one of the black holes. But there exist another set (of measure zero
but non-null dimension) of orbits which are not trapped by a single
black hole: the unstable periodic orbits in which the particle bounces
forever between different repulsion centers or, more precisely,
between different curves of zero velocity. This is the kind of
scenario in which chaotic scattering is likely to appear\cite{Ott}.
Although this system does not fit exactly in one of the classes in
which chaotic scattering is usually analyzed, i.e., gradient systems
and hard disks or spheres, we expect to have a similar qualitative
behavior because we still have curves of zero velocity (which are the
curves in which the energy equals the potential energy in the case
of gradient systems and the perimeter of hard disks). 

In the following, we consider a test particle with $e/m=2$ which is
sent from the point $(-4,b,0)$, for different values of the impact
parameter $b$, and with initial velocity $(v,0,0)$. The $v$ value is
selected in order to have always an energy per unit mass $E=-0.45$. As
can be seen in Fig.~\ref{fig:1}, for this value of the energy per unit
mass there exist a curve of zero velocity around each black hole
preventing orbits coming from infinity to be trapped by the black
holes. 

The system~(\ref{xdot})-(\ref{vdot}) is numerically solved by means of 
an ordinary-differential-equation solver\cite{ODE}. The quality of the 
numerical results is tested by using different integration schemes, 
ranging from the very stable embedded Runge-Kutta code of eight order 
due to Dormand and Prince to very fast extrapolation routines. All 
codes have adaptive step size control and we check that smaller 
tolerances do not change the results. Furthermore, the constancy of 
the energy~(\ref{energy}) is monitored to test the integration 
accuracy. 

In Fig.~\ref{fig:1} we see four orbits which correspond to very close 
initial conditions, but look very different after bouncing several 
times on different curves of zero velocity. This sensitive dependence 
on initial conditions is the hallmark of chaos. To explore this 
further 
we have chosen a large number of points ($\sim 2\times10^5$) in the 
interval $0.34 \le b \le 0.42$ and integrated the system for each 
initial condition until the particle escapes the system. (In practice 
we consider that the particle has abandoned the system when $|{\bf x}| 
> 10$, because it cannot return if $E<0$.) Then we compute the 
scattering angle $\theta$ between the velocity $\bf v$ and the $x$ 
axis. In Fig.~\ref{fig:2} we have plot 2000 points of the scattering 
function $\theta(b)$. In some ranges the function is continuous and 
the points are located on a smooth curve, but in other intervals 
the points 
are wildly scattered because $\theta(b)$ is discontinuous at the 
points of 
a Cantor set, as can be seen in the blowups of Figs.~\ref{fig:3} 
and~\ref{fig:4} where smaller and smaller subintervals are analyzed. 
The (approximate) scale invariance typical of fractals is apparent. 
This kind of phenomenon is called ``chaotic scattering''\cite{Ott}.

\section{Time Delays and the Uncertainty Dimension}
\label{sec:fractal}

Exactly as in ordinary Hamiltonian systems\cite{Ott}, the singularity 
points of the scattering function correspond to unstable periodic 
orbits in which the particles bounces forever between the curves of 
zero velocities of different black holes. This can be seen in 
Fig.~\ref{fig:5} where the time the particle spends in the scattering 
region is plotted against the impact parameter $b$. This time goes to 
infinity precisely at the singularity points of $\theta(b)$. It is 
easy to understand that if two orbits are initially very close but 
then come near one of the unstable periodic orbits, they will spend a 
very long time inside the scattering region and the accumulated 
defocusing effect of the successive bounces will yield rather 
different exit angles. 

Though the fractal nature of the scattering function is apparent from 
the approximate autosimilarity shown in Figs.~\ref{fig:2}, \ref{fig:3} 
and~\ref{fig:4}, we can make this fact more quantitative by using a 
variation of the ``uncertainty exponent technique''\cite{Ott}, which 
is also used in determining the dimension of fractal basin boundaries. 
We select a uncertainty value $\epsilon$ and sample the interval 
under study by choosing points in the form $b_{i+1}=b_i+\epsilon$. If 
for the three values $b_{i-1}$, $b_{i}$ and $b_{i+1}$ the particle 
scatters upward or the three values correspond to a downward 
scattering, we say that the $b_i$ value is $\epsilon$-certain. The 
remaining $b_i$ points are the $\epsilon$-uncertain values and 
correspond to cases in which an error of magnitude $\epsilon$ in the 
determination of the initial conditions will prevent us from 
predicting if the particle will finally escape the scattering region 
upward or downward. A measure of the uncertainty is thus given by the 
fraction of $\epsilon$-uncertain values which we will call 
$f(\epsilon)$. 

It has been found in many different contexts that this magnitude 
scale with respect to $\epsilon$ according to a power law:
 \begin{equation}
f(\epsilon)\sim \epsilon^{\alpha},
 \label{power}
 \end{equation}
where $\alpha$ is the ``uncertainty exponent.'' When the set of 
uncertain points (in the limit $\epsilon\to 0$) is not fractal it is 
easy to see that $\alpha = 1-D$, where $D$ is the ordinary dimension 
of the set. This relation is extended to fractal cases by defining
the ``uncertainty dimension'' as $D\equiv 1-\alpha$ and it has been 
conjectured that this definition will coincide with the ``box 
counting'' dimension\cite{Ott}. The meaning of the uncertainty 
dimension is that if we divide by 10 the error in the determination 
of 
the impact parameter, we only reduce by an amount of $10^{1-D}$ the 
error in the prediction of the late-time behavior. High values of $D$ 
($0<= D <1$ in our case) make very difficult improving our 
predictions. This is the obstacle to predictability in chaotic 
scattering. 

In Fig.~\ref{fig:6}, some values of the fraction of 
$\epsilon$-uncertain values $f(\epsilon)$ in the interval $0.34 \le b 
\le 0.42$ are plotted on a log-log scale. One sees that the points are 
fitted very well by a straight line of slope $\alpha=0.49$. We 
conclude that the power law~(\ref{power}) is satisfied and the 
uncertainty dimension of the singularity set of the system under study 
is 
 \begin{equation}
D=1-\alpha=0.51.
 \end{equation}
This dimension provides a coordinate independent measure of
chaos\cite{invariant}.

\section{Final Comments}
\label{sec:comments}

We have shown that a charged test particle may experience the
phenomenon of chaotic scattering around a static system formed by four
extreme Reissner-Nordstr\"om black holes: in some cases the scattering
angle (and time delay) function is singular at the points of a Cantor
set. As mentioned above, another manifestation of the existence of a
nonattracting chaotic set, fractal basin boundaries, was previously
known for MP solutions, but for other values of the parameters. In
some sense this kind of chaotic behavior is less surprising, because
it does not appear for $N=1, 2$ (which are the only cases with
an integrable Newtonian
limit\cite{Dettmann}). For $N > 2$, we have repeated the analysis
made above for the Newtonian limit, and it is not
difficult to find parameter ranges for which the scattering becomes
chaotic. Let us also mention that the same happens in the 
relativistic case. For instance, we have
found that chaotic scattering also happens around three black holes,
though we have chosen to present here the more symmetric case $N=4$.

 \acknowledgments

I would like to thank N.\ J.\ Cornish for useful criticisms and for
pointing out some relevant references.
This work has been supported by The University of the Basque Country
under contract UPV/EHU 172.310-EB036/95.

\newpage

\begin{figure} 
\caption{Energy per unit mass
(\protect\ref{energy}) for ${\bf v}=0$ and $e/m=2$.}
\label{fig:0} 
\end{figure}

\begin{figure} 
\caption{Four black holes at points $(\pm1,\pm1,0)$ have 
curves of zero velocity around them for $E=-0.45$
and $e/m=2$. Orbits corresponding to 
(1) $b=0.34975$, (2) $b=0.34971875$, 
(3) $b=0.3496875$ and (4) $b=0.34965625$ are also depicted.}
\label{fig:1} 
\end{figure}

\begin{figure} 
\caption{The scattering function $\theta(b)$ for 
$0.34 \le b \le 0.42$.}
\label{fig:2} 
\end{figure}

\begin{figure} 
\caption{A blowup of Fig.~(\protect\ref{fig:2}) showing the scattering 
function $\theta(b)$ for $0.36 \le b \le 0.44$.}
\label{fig:3} 
\end{figure}

\begin{figure} 
\caption{A further blowup of Figs.~(\protect\ref{fig:2}) 
and~(\protect\ref{fig:3}).}
\label{fig:4} 
\end{figure}

\begin{figure} 
\caption{The time-delay function indicating the time the particle 
spends in the scattering region for each value of the impact 
parameter. It becomes infinite at the singularity points of 
Fig.~(\protect\ref{fig:2}).} 
\label{fig:5} 
\end{figure}

\begin{figure} 
\caption{A log-log plot of the fraction of uncertain values 
$f(\epsilon)$ for different resolutions $\epsilon$. The straight line 
fit has a slope $\alpha=0.49$.} 
\label{fig:6} 
\end{figure}

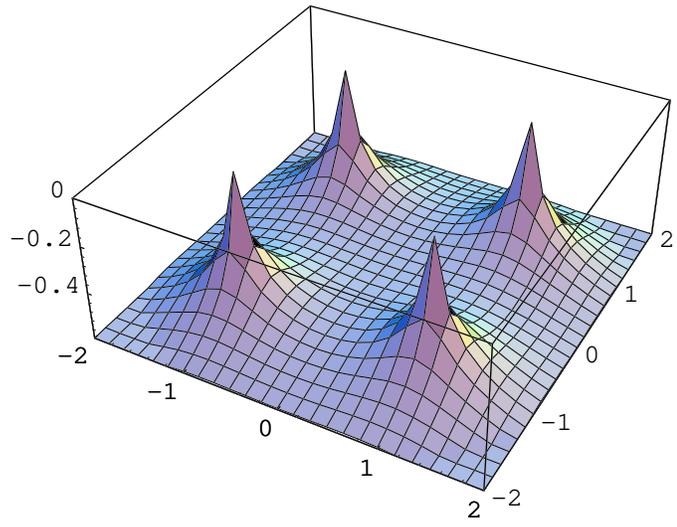
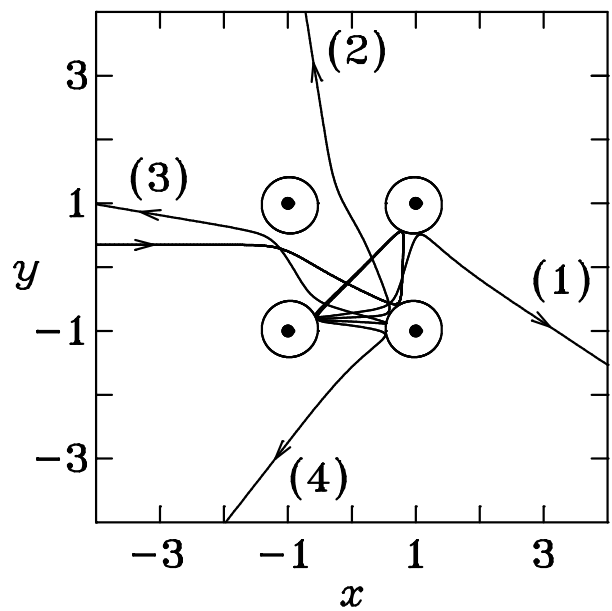
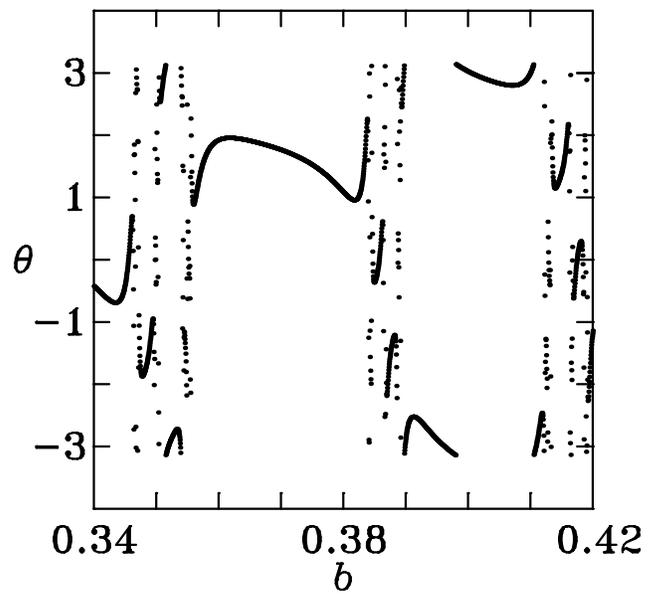
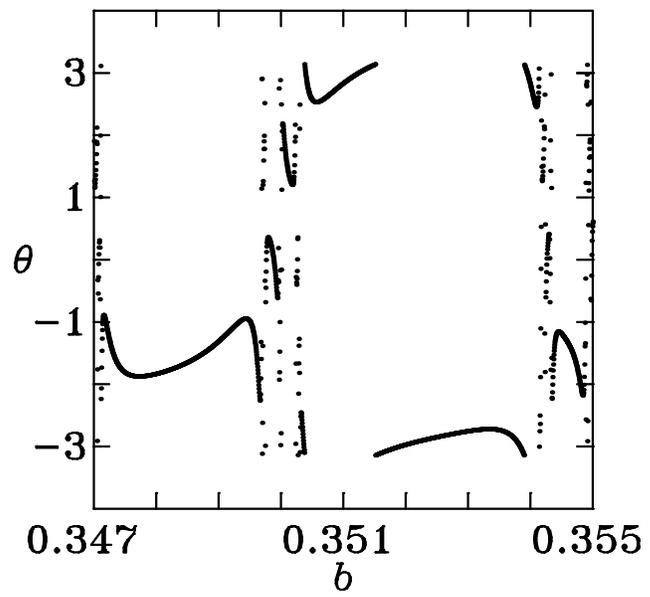
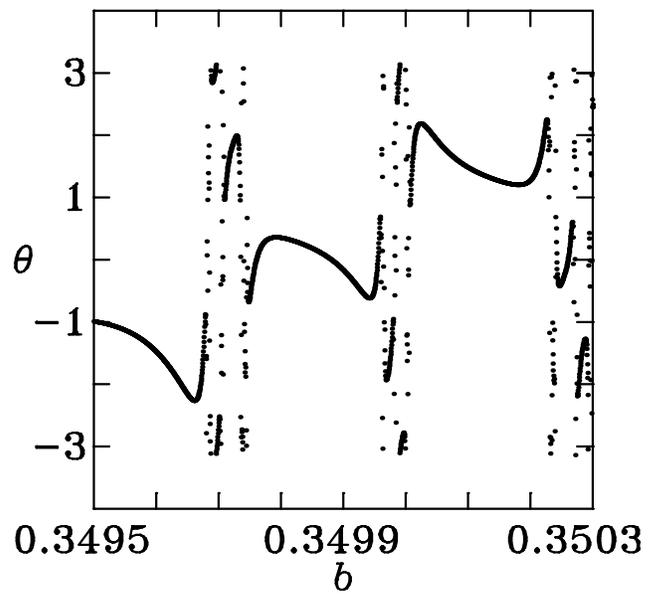
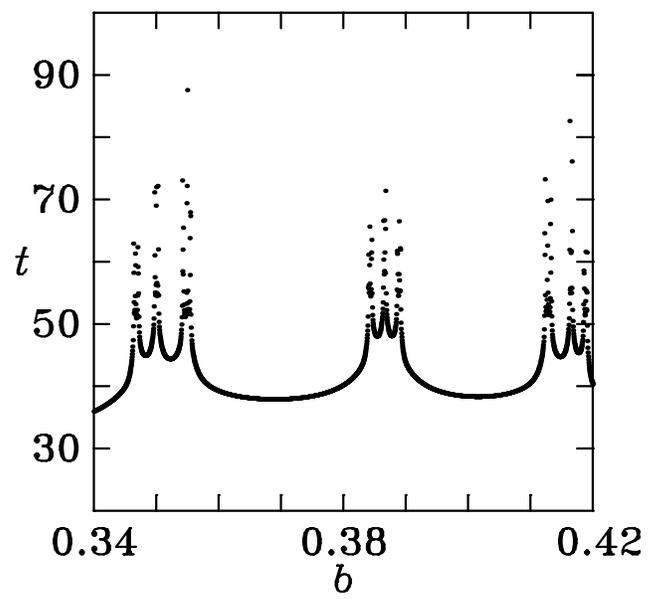
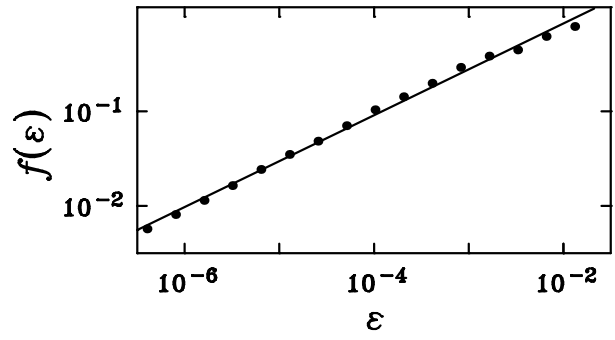
\end{document}